\documentclass{article}
\usepackage{times}
\usepackage{amsfonts}
\begin{document}
\title{Gerbes and Heisenberg's Uncertainty Principle}
\author{Jos\'e M. Isidro\\
Instituto de F\'{\i}sica Corpuscular (CSIC--UVEG)\\
Apartado de Correos 22085, Valencia 46071, Spain\\
{\tt jmisidro@ific.uv.es}}

\maketitle

\begin{abstract}

\noindent
We prove that a gerbe with a connection can be defined on classical phase space, taking the U(1)--valued phase of certain Feynman path integrals as \v Cech 2--cocycles. A quantisation condition on the corresponding 3--form field strength is proved to be equivalent to Heisenberg's uncertainty principle.
\end{abstract}
\noindent

\tableofcontents

\section{Introduction}\label{laagpttn}

{}Feynman's quantum--mechanical exponential of the classical action $S$,
\begin{equation}
\exp\left({\rm i}\frac{S}{\hbar}\right),
\label{bonita}
\end{equation}
has an interpretation in terms of {\it gerbes} \cite{GERBES}. The latter are geometrical structures developed recently, that have found interesting applications in several areas of geometry and theoretical physics \cite{RECENT}.  For the basics in the theory of gerbes the reader may want to consult the nice review \cite{PICKEN}.
We have in ref. \cite{GOLM} constructed a gerbe with a connection over the configuration space $\mathbb{F}$ corresponding to $d$ independent degrees of freedom. Specifically, the U(1)--valued phase of the quantum--mechanical transition amplitude $\langle q_2 t_2\vert q_1 t_1 \rangle$,
\begin{equation}
\frac{\langle q_2 t_2\vert q_1 t_1 \rangle}{\vert\langle q_2 t_2\vert q_1 t_1 \rangle\vert}=\exp\left({\rm i}\arg\, \langle q_2 t_2\vert q_1 t_1 \rangle\right),
\label{phph}
\end{equation}
is closely related to the trivialisation of a gerbe on $\mathbb{F}$. This fact can be used in order to prove that the semiclassical {\it vs.}\/ strong--quantum duality $S/\hbar\leftrightarrow\hbar/S$ of ref. \cite{ME} is equivalent to a Heisenberg--algebra noncommutativity \cite{NCG} for the space coordinates. The connection on the gerbe is interpreted physically as a {\it Neveu--Schwarz field}\/ $B_{\mu\nu}$ or, equivalently, as the magnetic background \cite{MAGNETIC} that causes space coordinates to stop being commutative and close a Heisenberg algebra instead.

Now the transition amplitude $\langle q_2 t_2\vert q_1 t_1 \rangle$ is proportional to the path integral 
\begin{equation}
\int{\rm D}q\,\exp\left(\frac{{\rm i}}{\hbar}\int_{t_1}^{t_2}{\rm d}t\,{\cal L}\right).
\label{kkeet}
\end{equation}
Whenever the Hamiltonian ${\cal H}(q,p)$ depends quadratically on $p$, eqn. (\ref{kkeet}) is the result of integrating over the momenta in the path integral
\begin{equation}
\int{\rm D}q\int{\rm D}p\,\exp\left(\frac{{\rm i}}{\hbar}\int_{t_1}^{t_2}{\rm d}t\left[p\dot q-{\cal H}(q,p)\right]\right).
\label{kklln}
\end{equation} 
In this sense the integral (\ref{kklln}) over phase space $\mathbb{P}$ is more general than the integral (\ref{kkeet}) over configuration space $\mathbb{F}$.

On the other hand, Heisenberg's uncertainty principle $\Delta Q\Delta P\geq \hbar/2$ can be derived from the Heisenberg algebra $[Q, P]={\rm i}\hbar$. In turn, the latter can be traced back to the corresponding classical Poisson brackets on $\mathbb{P}$. If, as shown in refs. \cite{GOLM, MAGNETIC}, a gerbe potential $B_{\mu\nu}$ on configuration space $\mathbb{F}$ is responsible for a Heisenberg algebra between space coordinates, then it makes sense to look for an interpretation of the uncertainty principle in terms of gerbes on classical phase space $\mathbb{P}$.

With this starting point, the purpose of this article is twofold:\\
{\it i)} To extend the formalism of ref. \cite{GOLM} from configuration space $\mathbb{F}$ to classical phase space $\mathbb{P}$, in order to construct a gerbe over the latter.\\
{\it ii)} To derive Heisenberg's uncertainty principle from the 3--form field strength on the above gerbe.

\section{A gerbe on classical phase space}\label{ptalg}

\subsection{The gerbe}\label{bbggr}

Classical phase space $\mathbb{P}$ is a $2d$--dimensional symplectic manifold endowed with the symplectic 2--form
\begin{equation}
\omega=\sum_{j=1}^d{\rm d}q^j\wedge{\rm d}p_j,
\label{ptkmd}
\end{equation}
when expressed in Darboux coordinates. The canonical 1--form $\theta$ on $\mathbb{P}$ defined as \cite{ARNOLD}
\begin{equation}
\theta:=-\sum_{j=1}^dp_j{\rm d}q^j
\label{ffrm}
\end{equation}
satisfies
\begin{equation}
{\rm d}\theta=\omega.
\label{gatita}
\end{equation}

Let $\{U_{\alpha}\}$ be a good cover of $\mathbb{P}$ by open sets $U_{\alpha}$. Pick any two points $(q_{\alpha_1},p_{\alpha_1})$ and $(q_{\alpha_2},p_{\alpha_2})$ on $\mathbb{P}$, respectively covered by the coordinate charts $U_{\alpha_1}$ and $U_{\alpha_2}$. The transition amplitude $\langle q_{\alpha_2} t_{\alpha_2}\vert q_{\alpha_1} t_{\alpha_1} \rangle$ is proportional to the path integral (\ref{kklln}):
\begin{equation}
\langle q_{\alpha_2} t_{\alpha_2}\vert q_{\alpha_1} t_{\alpha_1} \rangle\sim\int{\rm D}q\int{\rm D}p\,\exp\left(-\frac{{\rm i}}{\hbar}\int_{t_{\alpha_1}}^{t_{\alpha_2}}\left(\theta+{\cal H}{\rm d}t\right)\right).
\label{tobnn}
\end{equation}
The momenta $p$ being integrated over in (\ref{tobnn}) are unconstrained, while the coordinates $q$ satisfy the boundary conditions $q(t_{\alpha_j})=q_{\alpha_j}$ for $j=1,2$. Throughout, the $\sim$ sign will stand for {\it proportionality}\/: path integrals are defined up to some (usually divergent) normalisation. However all such normalisation factors cancel in the ratios of path integrals that we are interested in, such as (\ref{uuno}), (\ref{howw}) and (\ref{kkju}) below. The combination $\theta+{\cal H}{\rm d}t$, which we will denote by $\lambda$, is the integral invariant of Poincar\' e--Cartan \cite{ARNOLD}:
\begin{equation}
\lambda:=\theta+{\cal H}{\rm d}t.
\label{ji}
\end{equation}

Let $\mathbb{L}_{\alpha_1\alpha_2}\subset\mathbb{P}$ denote an oriented trajectory connecting $(q_{\alpha_1},p_{\alpha_1})$ to $(q_{\alpha_2},p_{\alpha_2})$. 
as time runs from $t_{\alpha_1}$ to $t_{\alpha_2}$. We define $\tilde a_{\alpha_1\alpha_2}$ as the following functional integral over all trajectories $\mathbb{L}_{\alpha_{1}\alpha_2}$ connecting $(q_{\alpha_1}, p_{\alpha_1})$ to $(q_{\alpha_2}, p_{\alpha_2})$:
\begin{equation}
\tilde a_{\alpha_1\alpha_2}\sim\int{\rm D}\mathbb{L}_{\alpha_{1}\alpha_2}\exp\left(-\frac{{\rm i}}{\hbar}\int_{\mathbb{L}_{\alpha_1\alpha_2}}\lambda\right).
\label{bbnmrk}
\end{equation}
The integral (\ref{bbnmrk}) differs from the transition amplitude (\ref{tobnn}) in that the momenta $p$ in the latter are unconstrained, while the momenta $p$ in (\ref{bbnmrk}) satisfy the same boundary conditions as the coordinates $q$. With this proviso we will continue to call $\tilde a_{\alpha_1\alpha_2}$ a {\it probability amplitude}. Its U(1)--valued phase is
\begin{equation}
a_{\alpha_1\alpha_2}:=\frac{\tilde a_{\alpha_1\alpha_2}}{\vert\tilde a_{\alpha_1\alpha_2}\vert}.
\label{uuno}
\end{equation}
Next assume that $U_{\alpha_1}\cap U_{\alpha_2}$ is nonempty,
\begin{equation}
U_{\alpha_{1}\alpha_2}:=U_{\alpha_1}\cap U_{\alpha_2}\neq\phi,
\label{lkjh}
\end{equation}
and let $(q_{\alpha_{12}},p_{\alpha_{12}})\in U_{\alpha_1\alpha_2}$. For $\alpha_1$ and $\alpha_2$ fixed we define 
$$
\tau_{\alpha_1\alpha_2}\colon U_{\alpha_{1}\alpha_2}\longrightarrow {\rm U(1)}
$$
\begin{equation}
\tau_{\alpha_1\alpha_2}(q_{\alpha_{12}},p_{\alpha_{12}}):=a_{\alpha_1\alpha_{12}} a_{\alpha_{12}\alpha_2}.
\label{bbdd}
\end{equation}
Thus $ \tau_{\alpha_1\alpha_2}(q_{\alpha_{12}},p_{\alpha_{12}})$ equals the U(1)--valued phase of the probability amplitude $\tilde a$ for the particle to start at $(q_{\alpha_1},p_{\alpha_1})\in U_{\alpha_1}$, then pass through $(q_{\alpha_{12}},p_{\alpha_{12}})\in U_{\alpha_{1}\alpha_2}$, and finally end at $(q_{\alpha_2},p_{\alpha_2})\in U_{\alpha_2}$. 
One readily verifies that (\ref{bbdd}) defines a gerbe trivialisation. We may rewrite the trivialisation (\ref{bbdd}) as
\begin{equation}
\tau_{\alpha_1\alpha_2}=\frac{\tilde\tau_{\alpha_1\alpha_2}}{\vert \tilde\tau_{\alpha_1\alpha_2}\vert}
\label{howw}
\end{equation}
where $\tilde\tau_{\alpha_1\alpha_2}$ is defined as the path integral
\begin{equation}
\tilde\tau_{\alpha_1\alpha_2}\sim\int{\rm D}\mathbb{L}_{\alpha_1\alpha_1}(\alpha_{12})\exp\left(-\frac{{\rm i}}{\hbar}\int_{\mathbb{L}_{\alpha_1\alpha_2}(\alpha_{12})}\lambda\right).
\label{efer}
\end{equation}
In $\tilde\tau_{\alpha_1\alpha_2}$ one integrates over all trajectories that, connecting $\alpha_1$ to $\alpha_2$, pass through the variable midpoint $\alpha_{12}$; the notation $\mathbb{L}_{\alpha_1\alpha_1}(\alpha_{12})$ stresses this fact. Therefore $\tilde\tau_{\alpha_1\alpha_2}$, and hence also $\tau_{\alpha_1\alpha_2}$, is a function on $U_{\alpha_1\alpha_2}$.

Next consider three points 
\begin{equation}
(q_{\alpha_1},p_{\alpha_1})\in U_{\alpha_1}, \qquad (q_{\alpha_2},p_{\alpha_2})\in U_{\alpha_2}, \qquad (q_{\alpha_3}, p_{\alpha_3})\in U_{\alpha_3}
\label{poiie}
\end{equation}
such that the triple overlap $U_{\alpha_1}\cap U_{\alpha_2}\cap U_{\alpha_3}$ is nonempty,
\begin{equation}
U_{\alpha_1\alpha_2\alpha_3}:=U_{\alpha_1}\cap U_{\alpha_2}\cap U_{\alpha_3}\neq\phi.
\label{lkvvjh}
\end{equation}
Once the trivialisation (\ref{bbdd}) is known, the 2--cocycle $g_{\alpha_1\alpha_2\alpha_3}$ defining a gerbe on $\mathbb{P}$ is given by \cite{GERBES}
$$
g_{\alpha_1\alpha_2\alpha_3}\colon U_{\alpha_1\alpha_2\alpha_3}\longrightarrow {\rm U(1)}
$$
$$
g_{\alpha_1\alpha_2\alpha_3}(q_{\alpha_{123}}, p_{\alpha_{123}}):=
$$
\begin{equation}
\tau_{\alpha_1\alpha_2}(q_{\alpha_{123}}, p_{\alpha_{123}})\tau_{\alpha_2\alpha_3}(q_{\alpha_{123}}, p_{\alpha_{123}})\tau_{\alpha_3\alpha_1}(q_{\alpha_{123}}, p_{\alpha_{123}}),
\label{asedcc}
\end{equation}
where all three $\tau$'s on the right--hand side are, by definition, evaluated at the same variable midpoint 
\begin{equation}
(q_{\alpha_{123}}, p_{\alpha_{123}})\in U_{\alpha_1\alpha_2\alpha_3}.
\label{ddmmd}
\end{equation}
In this way, $g_{\alpha_1\alpha_2\alpha_3}(q_{\alpha_{123}}, p_{\alpha_{123}})$ equals the U(1)--phase of the probability amplitude $\tilde a$ for the following transition (see figure\footnote{Figure available upon request.}): starting at $(q_{\alpha_1},p_{\alpha_1})$ we pass through $(q_{\alpha_{123}}, p_{\alpha_{123}})$ on our way to $(q_{\alpha_2},p_{\alpha_2})$; from here we cross $(q_{\alpha_{123}}, p_{\alpha_{123}})$ again on our way to $(q_{\alpha_3},p_{\alpha_3})$; finally from $(q_{\alpha_3},p_{\alpha_3})$ we once more pass through $(q_{\alpha_{123}}, p_{\alpha_{123}})$ on our way back to $(q_{\alpha_1},p_{\alpha_1})$. The complete closed trajectory is
\begin{equation}
\mathbb{L}_{\alpha_1\alpha_2\alpha_3}(\alpha_{123}):=\mathbb{L}_{\alpha_1\alpha_2}(\alpha_{123})+\mathbb{L}_{\alpha_2\alpha_3}(\alpha_{123})+\mathbb{L}_{\alpha_3\alpha_1}(\alpha_{123}).
\label{dfg}
\end{equation}
Being U(1)--valued, we can write $g_{\alpha_1\alpha_2\alpha_3}$ in eqn. (\ref{asedcc}) as the quotient 
\begin{equation}
g_{\alpha_1\alpha_2\alpha_3}=\frac{\tilde g_{\alpha_1\alpha_2\alpha_3}}{\vert\tilde g_{\alpha_1\alpha_2\alpha_3}\vert},
\label{kkju}
\end{equation}
where
\begin{equation}
\tilde g_{\alpha_1\alpha_2\alpha_3}\sim\int{\rm D}\mathbb{L}_{\alpha_1\alpha_2\alpha_3}(\alpha_{123})\exp\left(-\frac{{\rm i}}{\hbar}\int_{\mathbb{L}_{\alpha_1\alpha_2\alpha_3}(\alpha_{123})}\lambda\right).
\label{mrnna}
\end{equation}
This functional integral extends over all trajectories described in (\ref{dfg}). The notation $\mathbb{L}_{\alpha_1\alpha_2\alpha_3}(\alpha_{123})$ stresses the fact that all such paths traverse the variable midpoint $(q_{\alpha_{123}}, p_{\alpha_{123}})$. Therefore $\tilde g_{\alpha_1\alpha_2\alpha_3}$, and hece also $g_{\alpha_1\alpha_2\alpha_3}$, is a function on $U_{\alpha_1\alpha_2\alpha_3}$.

Consider now the first half of the leg $\mathbb{L}_{\alpha_1\alpha_2}(\alpha_{123})$, denoted $\frac{1}{2}\mathbb{L}_{\alpha_1\alpha_2}(\alpha_{123})$. The latter 
runs from $\alpha_1$ to $\alpha_{123}$. Consider also the second half of the leg $\mathbb{L}_{\alpha_3\alpha_1}(\alpha_{123})$, denoted $\frac{1}{2'}\mathbb{L}_{\alpha_3\alpha_1}(\alpha_{123})$, with a prime to remind us that it is the {\it second}\/ half: it runs back from $\alpha_{123}$ to $\alpha_1$ (see figure). The sum of these two half legs,
\begin{equation}
\frac{1}{2}\mathbb{L}_{\alpha_1\alpha_2}(\alpha_{123})+\frac{1}{2'}\mathbb{L}_{\alpha_3\alpha_1}(\alpha_{123}),
\label{mezzo}
\end{equation}
completes one roundtrip and it will, as a rule, enclose an area $\mathbb{S}_{\alpha_1}(\alpha_{123})$, unless the path from $\alpha_{123}$ to $\alpha_1$  happens to coincide exactly with the path from $\alpha_1$ to $\alpha_{123}$:
\begin{equation}
\partial\mathbb{S}_{\alpha_1}(\alpha_{123})=\frac{1}{2}\mathbb{L}_{\alpha_1\alpha_2}(\alpha_{123})+\frac{1}{2'}\mathbb{L}_{\alpha_3\alpha_1}(\alpha_{123}).
\label{soprano}
\end{equation}
Although the surface $\mathbb{S}_{\alpha_1}(\alpha_{123})$  is not unique, for the moment any such surface will serve our purposes. Analogous considerations apply to the other half legs $\frac{1}{2'}\mathbb{L}_{\alpha_1\alpha_2}(\alpha_{123})$, $\frac{1}{2}\mathbb{L}_{\alpha_3\alpha_1}(\alpha_{123})$, $\frac{1}{2}\mathbb{L}_{\alpha_2\alpha_3}(\alpha_{123})$ and $\frac{1}{2'}\mathbb{L}_{\alpha_2\alpha_3}(\alpha_{123})$ under cyclic permutations of 1,2,3 in the \v Cech indices $\alpha_1$, $\alpha_2$ and $\alpha_3$:
\begin{equation}
\partial\mathbb{S}_{\alpha_2}(\alpha_{123})=\frac{1}{2}\mathbb{L}_{\alpha_2\alpha_3}(\alpha_{123})+\frac{1}{2'}\mathbb{L}_{\alpha_1\alpha_2}(\alpha_{123}),
\label{tenor}
\end{equation}
\begin{equation}
\partial\mathbb{S}_{\alpha_3}(\alpha_{123})=\frac{1}{2}\mathbb{L}_{\alpha_3\alpha_1}(\alpha_{123})+\frac{1}{2'}\mathbb{L}_{\alpha_2\alpha_3}(\alpha_{123}).
\label{bajo}
\end{equation}
The boundaries of the three surfaces $\mathbb{S}_{\alpha_1}(\alpha_{123})$, $\mathbb{S}_{\alpha_2}(\alpha_{123})$ and $\mathbb{S}_{\alpha_3}(\alpha_{123})$ all pass through the variable midpoint (\ref{ddmmd}). We define their connected sum
\begin{equation}
\mathbb{S}_{\alpha_1\alpha_2\alpha_3}:=\mathbb{S}_{\alpha_1}+\mathbb{S}_{\alpha_2}+\mathbb{S}_{\alpha_3}.
\label{wdvdd}
\end{equation}
In this way we have
\begin{equation}
\mathbb{L}_{\alpha_1\alpha_2\alpha_3}=\partial\mathbb{S}_{\alpha_1\alpha_2\alpha_3}=
\partial\mathbb{S}_{\alpha_1}+\partial\mathbb{S}_{\alpha_2}+\partial\mathbb{S}_{\alpha_3}.
\label{bbvvdd}
\end{equation}
It must be borne in mind that $\mathbb{L}_{\alpha_1\alpha_2\alpha_3}$ is a function of the variable midpoint $\alpha_{123}\in U_{\alpha_1\alpha_2\alpha_3}$, even if we no longer indicate this explicitly.  Eventually one, two or perhaps all three of $\mathbb{S}_{\alpha_1}$, $\mathbb{S}_{\alpha_2}$ and $\mathbb{S}_{\alpha_3}$ may degenerate to a curve connecting the midpoint (\ref{ddmmd}) with $\alpha_1$, $\alpha_2$ or $\alpha_3$, respectively. Whenever such is the case for all three surfaces, the closed trajectory $\mathbb{L}_{\alpha_1\alpha_2\alpha_3}$ cannot be expressed as the boundary of a 2--dimensional surface $\mathbb{S}_{\alpha_1\alpha_2\alpha_3}$. In what follows we will however exclude this latter possibility, so that at least one of the three surfaces on the right--hand side of (\ref{wdvdd}) does not degenerate to a curve.

One further comment is in order. The gerbe we have constructed is defined on phase space $\mathbb{P}$. If $\mathbb{R}$ denotes the time axis, we have the natural inclusion $\iota\colon\mathbb{P}\rightarrow\mathbb{P}\times\mathbb{R}$. The 1--form $\lambda$ is defined on $\mathbb{P}\times\mathbb{R}$, but all the line integrals we have considered here in fact involve its pullback $\iota^*\lambda$ to $\mathbb{P}$, rather than $\lambda$ itself, even if this has not been denoted explicitly. Moreover, the term ${\cal H}{\rm d}t$ within $\lambda$ will drop out of our calculations, as we will see in section \ref{ppoyy}. An equivalent statement of this fact is that we are working on $\mathbb{P}$ at a fixed value of the time.

\subsection{The steepest--descent approximation to the 2--cocycle}\label{kkbbww}

We can approximate the path--integral (\ref{mrnna}) by the method of steepest descent \cite{ZJ}. We are given a path integral
\begin{equation}
\int{\rm D}f\exp\left({\cal F}[f]\right),
\label{rizz}
\end{equation}
where the argument of the exponential contains a 1--dimensional integral
\begin{equation}
{\cal F}[f]:=\int{\rm d}t\, f\left(u_i(t), \dot u_i(t),  t\right),\quad i=1,\ldots r.
\label{tin}
\end{equation}
Consider the diagonal $r\times r$ matrix $M$ whose $i$--th entry $m_i$ equals
\begin{equation}
m_i:= \frac{\partial^2f}{\partial\dot u_i^2},\qquad i=1,\ldots, r.
\label{hessian}
\end{equation}
If the extremals $u^{(0)}_i$, $i=1,\ldots,r$, make the integral ${\cal F}$ a minimum, then all the $m_i$, evaluated at the extremals $u^{(0)}_i$, are nonnegative \cite{CH}. Hence
\begin{equation}
{\rm det}M^{(0)}=\prod_{i=1}^rm^{(0)}_i\geq 0,
\label{xets}
\end{equation}
the superindex ${}^{(0)}$ standing for ``evaluation at the extremal". We will assume that det$M^{(0)}>0$. Then the steepest descent approximation to (\ref{rizz}) yields
\begin{equation}
\int{\rm D}f\exp\left({\cal F}[f]\right)\sim\left(-{\rm det}M^{(0)}\right)^{-1/2}\,\exp\left({\cal F}[f^{(0)}]\right).
\label{rizzw}
\end{equation}

In our case (\ref{mrnna}), the saddle point is given by those closed paths $\mathbb{L}_{\alpha_1\alpha_2\alpha_3}^{(0)}$ that minimise the integral
\begin{equation}
\int_{\mathbb{L}_{\alpha_1\alpha_2\alpha_3}}\lambda
\label{reaa}
\end{equation}
for fixed $\alpha_1$, $\alpha_2$ and $\alpha_3$. The $u_i(t)$ of eqns. (\ref{tin})--(\ref{rizzw}) are replaced by the pullbacks $q_j(t)$, $p^j(t)$, to the path $\mathbb{L}_{\alpha_1\alpha_2\alpha_3}$, of the Darboux coordinates $q_j$, $p^j$ on phase space $\mathbb{P}$. In particular we have $r=2d$. Altogether, the steepest descent approximation (\ref{rizzw}) to the path integral (\ref{mrnna}) leads to
\begin{equation}
\tilde g_{\alpha_1\alpha_2\alpha_3}^{(0)}\sim\left(\frac{{\rm i}}{\hbar}\,{\rm det}M^{(0)}\right)^{-1/2}\exp\left(-\frac{{\rm i}}{\hbar}\int_{\mathbb{L}_{\alpha_1\alpha_2\alpha_3}^{(0)}}\lambda\right).
\label{menospeor}
\end{equation} 
Now det$M^{(0)}>0$ so, by eqn. (\ref{kkju}), it does not contribute to the 2--cocycle. After dropping an irrelevant ${\rm e}^{-{\rm i}\pi/{4}}$ we finally obtain
\begin{equation}
g_{\alpha_1\alpha_2\alpha_3}^{(0)}=\exp\left(-\frac{{\rm i}}{\hbar}\int_{\mathbb{L}_{\alpha_1\alpha_2\alpha_3}^{(0)}}\lambda\right).
\label{yya}
\end{equation}
Eqn. (\ref{yya}) gives the steepest--descent approximation $g_{\alpha_1\alpha_2\alpha_3}^{(0)}$ to the 2--cocycle $g_{\alpha_1\alpha_2\alpha_3}$ defining the gerbe on phase space $\mathbb{P}$. As already remarked before eqn. (\ref{bbvvdd}),  $g_{\alpha_1\alpha_2\alpha_3}^{(0)}$ is a function of the variable midpoint (\ref{ddmmd}) through the integration path $\mathbb{L}_{\alpha_1\alpha_2\alpha_3}^{(0)}$, even if we no longer indicate this explicitly.

\subsection{The connection}\label{knnk}

On a gerbe determined by the 2--cocycle $g_{\alpha_1\alpha_2\alpha_3}$, a connection is specified by forms $A, B, H$ satisfying  \cite{GERBES}
\begin{eqnarray}
H\vert_{U_{\alpha}}&=&{\rm d}B_{\alpha}\\
B_{\alpha_2}-B_{\alpha_1}&=&{\rm d}A_{\alpha_1\alpha_2}\\
A_{\alpha_1\alpha_2}+A_{\alpha_2\alpha_3}+A_{\alpha_3\alpha_1}&=&g^{-1}_{\alpha_1\alpha_2\alpha_3}{\rm d}g_{\alpha_1\alpha_2\alpha_3}.
\label{ktpyy}
\end{eqnarray}
The gerbe is called {\it flat}\/ if $H=0$.

We can use eqn. (\ref{yya}) in order to compute the connection, at least to the same order of accuracy as the 2--cocycle $g_{\alpha_1\alpha_2\alpha_3}$ itself:
\begin{equation}
A_{\alpha_1\alpha_2}^{(0)}+A_{\alpha_2\alpha_3}^{(0)}+A_{\alpha_3\alpha_1}^{(0)}=\left(g^{(0)}_{\alpha_1\alpha_2\alpha_3}\right)^{-1}{\rm d}g_{\alpha_1\alpha_2\alpha_3}^{(0)}.
\label{accx}
\end{equation}
We will henceforth drop the superindex ${}^{(0)}$, with the understanding that all our computations have been done in the steepest--descent approximation.
We find
\begin{equation}
A_{\alpha_1\alpha_2}=-\frac{{\rm i}}{\hbar}\lambda_{\alpha_1\alpha_2}=-\frac{{\rm i}}{\hbar}\left(\theta+{\cal H}{\rm d}t\right)_{\alpha_1\alpha_2}.
\label{dinf}
\end{equation}
Therefore
\begin{equation}
B_{\alpha_2}-B_{\alpha_1}={\rm d}A_{\alpha_1\alpha_2}=-\frac{{\rm i}}{\hbar}\left(\omega+{\rm d}{\cal H}\wedge{\rm d}t\right)_{\alpha_1\alpha_2}.
\label{rrtt}
\end{equation}
On constant--energy submanifolds of phase space the above simplifies to
\begin{equation}
B_{\alpha_2}-B_{\alpha_1}=-\frac{{\rm i}}{\hbar}\,\omega_{\alpha_1\alpha_2}.
\label{ffwe}
\end{equation}
We will henceforth assume that we are working on constant--energy submanifolds of phase space.

\subsection{Symplectic area}\label{ppoyy}

Let $\mathbb{S}\subset\mathbb{P}$ be a 2--dimensional surface with the boundary $\partial\mathbb{S}=\mathbb{L}$. By Stokes' theorem and eqns. (\ref{gatita}), (\ref{ji}),
\begin{equation}
\int_{\mathbb{L}}\lambda=\int_{\partial\mathbb{S}}\lambda
=\int_{\mathbb{S}}{\rm d}\lambda=\int_{\mathbb{S}}\left(\omega+{\rm d}{\cal H}\wedge{\rm d}t\right).
\label{llkdes}
\end{equation}
Let us pick $\mathbb{S}$ such that it is a constant--energy surface, or else a constant--time surface. Then
\begin{equation}
\int_{\mathbb{L}}\lambda=\int_{\mathbb{S}}\omega.
\label{lxxlkdes}
\end{equation}
The right--hand side of eqn. (\ref{lxxlkdes}) does not depend on the particular surface $\mathbb{S}$ chosen because  \cite{ARNOLD}
\begin{equation}
{\rm d}\omega=0.
\label{schuso}
\end{equation}
Next pick $\mathbb{S}$ as $\mathbb{S}_{\alpha_1\alpha_2\alpha_3}$ in eqn. (\ref{wdvdd}). By eqn. (\ref{lxxlkdes}), the 2--cocycle  (\ref{yya}) reads
\begin{equation}
g_{\alpha_1\alpha_2\alpha_3}=\exp\left(-\frac{{\rm i}}{\hbar}\int_{\mathbb{S}_{\alpha_1\alpha_2\alpha_3}}\omega\right).
\label{yyaj}
\end{equation}
The above can be given a nice quantum--mechanical interpretation. The integral 
\begin{equation}
\frac{1}{\hbar}\int_{\mathbb{S}_{\alpha_1\alpha_2\alpha_3}}\omega
\label{laav}
\end{equation}
equals the symplectic area of the surface $\mathbb{S}_{\alpha_1\alpha_2\alpha_3}$ in units of $\hbar$. In the WKB approximation \cite{ZJ}, the absolute value of (\ref{laav}) is proportional to the number of quantum states contributed by the surface $\mathbb{S}_{\alpha_1\alpha_2\alpha_3}$ to the Hilbert space of quantum states. Now the steepest descent approximation used here is a rephrasing of the WKB method. We conclude that {\it the 2--cocycle $g_{\alpha_1\alpha_2\alpha_3}$ equals the exponential of ($-{\rm i}$ times) the number of quantum states contributed by any surface $\mathbb{S}_{\alpha_1\alpha_2\alpha_3}$ bounded by the closed loop $\mathbb{L}_{\alpha_1\alpha_2\alpha_3}$.} The constant--energy condition on the surface translates quantum--mechanically into the stationarity of the corresponding states. The steepest--descent approximation minimises the symplectic area of the open, constant--energy surface $\mathbb{S}_{\alpha_1\alpha_2\alpha_3}$.

\subsection{The field strength}\label{fgord}

By eqns. (\ref{ffwe}) and (\ref{schuso}) it follows that ${\rm d}B_{\alpha_1}={\rm d}B_{\alpha_2}$. This implies that the 3--form field strength $H$, contrary to the 2--form potential $B$, is globally defined on $\mathbb{P}$. Consider now a 3--dimensional volume $\mathbb{V}\subset\mathbb{P}$ whose boundary is a 2--dimensional closed surface $\mathbb{S}$. If $\mathbb{V}$ is connected and simply connected we may, without loss of generality, take $\mathbb{V}$ to be a solid ball, so $\mathbb{S}=\partial\mathbb{V}$ is a sphere. Let us cover $\mathbb{S}$ by stereographic projection.  This gives us two coordinate charts, respectively centred around the north and south poles on the sphere. Each chart is diffeomorphic to a copy of the plane $\mathbb{R}^2$. Each plane covers the whole $\mathbb{S}$ with the exception of the opposite pole. The intersection of these two charts is the whole sphere $\mathbb{S}$ punctured at its north and south poles. The situation just described is perfect for a discussion of eqn.  (\ref{ffwe}). Let us embed the chart $\mathbb{R}^2_{\alpha_1}$ centred at the north pole within the open set $ U_{\alpha_1}$, {\it i.e.}, $\mathbb{R}^2_{\alpha_1}\subset U_{\alpha_1}$, if necessary by means of some diffeomorphism. Analogously, for the south pole we have $\mathbb{R}^2_{\alpha_2}\subset U_{\alpha_2}$. There is also no loss of generality in assuming that only two points on the sphere $\mathbb{S}$ (the north and south poles) remain outside the 2--fold overlap $U_{\alpha_1}\cap U_{\alpha_2}$. By Stokes' theorem,
\begin{equation}
\int_{\mathbb{V}}H=\int_{\mathbb{V}}{\rm d}B=\int_{\partial\mathbb{V}}B=\int_{\mathbb{S}}B=\int_{\mathbb{R}^2_{\alpha_2}}B-\int_{\mathbb{R}^2_{\alpha_1}}B,
\label{kote}
\end{equation}
and, by eqn. (\ref{ffwe}),
\begin{equation}
\int_{\mathbb{V}}H=-\frac{{\rm i}}{\hbar}\int_{\mathbb{R}^2-\{0\}}\omega,
\label{kabaka}
\end{equation}
where $\mathbb{R}^2-\{0\}$ denotes either one of our two charts, punctured at its corresponding origin. Since $\mathbb{R}^2-\{0\}$ falls short of covering the whole sphere $\mathbb{S}$ by just two points (the north and south poles), and the latter have zero measure, we may just as well write
\begin{equation}
\int_{\mathbb{V}}H=-\frac{{\rm i}}{\hbar}\int_{\mathbb{S}}\omega.
\label{kakaka}
\end{equation}

Eqn. (\ref{kakaka}) is analogous to the Gauss law in electrostatics, with $H$ replacing the electric charge density 3--form and $\omega/{\rm i}\hbar$ replacing the corresponding surface flux 2--form. If our gerbe is nonflat, then $H$ may be regarded as a source term for the quantum states arising from a nonvanishing flux of $\omega/{\rm i}\hbar$ across the closed surface $\mathbb{S}$. On the contrary, the gerbe is flat if and only if every closed surface $\mathbb{S}\subset\mathbb{P}$ (satisfying the above requirements concerning $\mathbb{V}$) contributes no quantum states at all to the Hilbert space. This is tantamount to the statement that every closed surface $\mathbb{S}\subset\mathbb{P}$ (satisfying the above requirements concerning $\mathbb{V}$) has zero symplectic area. In other words, the gerbe is flat if, and only if,  {\it open}\/ surfaces $\mathbb{S}$ are the {\it unique}\/ sources of quantum states. Then the mechanism responsible for the generation of quantum states is a nonvanishing symplectic area of the open surface $\mathbb{S}$. Equivalently, by eqn. (\ref{lxxlkdes}), this mechanism is a nonvanishing circulation of the Poincar\'e--Cartan 1--form $\lambda$ along its boundary $\mathbb{L}$.

Now Heisenberg's uncertainty principle implies a discretisation, or quantisation, of symplectic area in units of $\hbar$. To begin with let us consider {\it closed}\/ surfaces $\mathbb{S}$ inside phase space. Then, within the WKB approximation \cite{ZJ},
\begin{equation}
\frac{1}{\hbar}\int_{\mathbb{S}}\omega=2\pi n, \qquad n\in\mathbb{Z},\qquad \partial\mathbb{S}=0,
\label{pauli}
\end{equation}
which, by eqn. (\ref{kakaka}), is tantamount to quantising the volume integral of $H/2\pi {\rm i}$. In turn, this can be recast as the quantisation condition \cite{GERBES, ORLANDO}
\begin{equation}
\frac{1}{2\pi{\rm i}}\int_{\mathbb{V}}H\in\mathbb{Z}, \qquad \partial\mathbb{V}=\mathbb{S},
\label{eqqwe}
\end{equation}
for all 3--dimensional, connected and simply connected volumes $\mathbb{V}\subset\mathbb{P}$. Starting from Heisenberg's principle we have obtained the quantisation condition (\ref{eqqwe}). Conversely, assume taking (\ref{eqqwe}) above as our starting point on phase space, and let us derive Heisenberg's principle. Given a 3--dimensional volume $\mathbb{V}\subset\mathbb{P}$ such that $\partial\mathbb{V}=\mathbb{S}$, eqns. (\ref{kakaka}) and (\ref{eqqwe}) imply that symplectic area is quantised on closed surfaces. This is an equivalent rendering of the uncertainty principle, at least on closed surfaces. 

Now {\it open}\/ surfaces within phase space have their symplectic area quantised according to the WKB rule \cite{ZJ}
\begin{equation}
\frac{1}{\hbar}\int_{\mathbb{S}}\omega=2\pi \left(n+\frac{1}{2}\right), \qquad n\in\mathbb{Z}, \qquad \partial\mathbb{S}\neq 0.
\label{paulix}
\end{equation}
Notice the additional $1/2$ in (\ref{paulix}) (open surfaces) as opposed to (\ref{pauli}) (closed surfaces). Consider now two open surfaces $\mathbb{S}_{1}$ and $\mathbb{S}_{2}$ such that $\partial\mathbb{S}_{1}=-\partial\mathbb{S}_{2}$. We can glue them along their common boundary to produce a closed surface to which the quantisation condition (\ref{pauli}) applies, hence (\ref{eqqwe}) follows. Conversely, if we start off from a gerbe on $\mathbb{P}$ satisfying eqn. (\ref{eqqwe}), let us prove that symplectic area is quantised on open surfaces as well. Consider a fixed open surface $\mathbb{S}_{1}$, plus a family of open surfaces $\mathbb{S}_{2}^{(\iota)}$ parametrised by a certain index $\iota$, such that $\partial\mathbb{S}_{2}^{(\iota)}=-\mathbb{S}_{1}$ for all $\iota$. Glue each $\mathbb{S}_{2}^{(\iota)}$ on to $\mathbb{S}_{1}$ along the common boundary, in order to obtain a family of closed surfaces $\mathbb{S}^{(\iota)}$. Symplectic area is quantised on all of the latter. Now $\mathbb{S}_{1}$ is fixed while the $\mathbb{S}_{2}^{(\iota)}$ are varied. As the index $\iota$ is arbitrary, the variations in the $\mathbb{S}_{2}^{(\iota)}$, hence in the $\mathbb{S}^{(\iota)}$, can be made arbitrary. Meanwhile the symplectic area of the $\mathbb{S}^{(\iota)}$, which is the sum of the areas of $\mathbb{S}_{1}$ and  $\mathbb{S}_{2}^{(\iota)}$, remains quantised as per eqn. (\ref{pauli}). This can only be the case if symplectic area is quantised on {\it open}\/ surfaces as well. Strictly speaking, this argument only establishes that the symplectic area of open surfaces is quantised as $2\pi k$, where $2k\in\mathbb{Z}$. The additional $1/2$ present in (\ref{paulix}) follows when $k\notin\mathbb{Z}$.

To summarise, {\it eqn. (\ref{eqqwe}) is an equivalent statement of Heisenberg's uncertainty principle}.

\section{Outlook}\label{kool}

A number of challenging questions arise.

We have worked in the WKB approximation; it would be interesting to compute higher quantum corrections to our results. Such corrections will generally depend on the dynamics. In this respect one could consider the approach of ref. \cite{MATONE}, where Planck's constant $\hbar$ is regarded as a dynamically--generated quantum scale. What modifications of the uncertainty principle this may bring about in our setup remains to be clarified. Current field--theoretic and string models certainly do lead to such modifications.

According to conventional folklore, ``the uncertainty principle prohibits quantum mechanics on phase space". Here have shown that endowing phase space with a gerbe and a connection is a way of quantising classical mechanics. In fact, phase space is becoming increasingly popular as a natural arena for quantum mechanics \cite{GOSSON}. 
Our conclusions also contribute towards a modern geometric view of quantum mechanics,  a beautiful presentation of which has been given in ref. \cite{GMS}. Last but not least, the ideas explored here are connected, not as remotely as it may on first sight appear, with quantum theories of gravity \cite{PERSPECTIVE}.

{\bf Acknowledgements}

It is a great pleasure to thank Albert--Einstein--Institut f\"ur Gravitationsphysik (Potsdam, Germany) for hospitality during the preparation of this article. This work has been supported by Ministerio de Educaci\'{o}n y Ciencia (Spain) through grant FIS2005--02761, by Generalitat Valenciana, by EU FEDER funds, by EU network MRTN--CT--2004--005104 ({\it Constituents, Fundamental Forces and Symmetries of the Universe}), and by Deutsche Forschungsgemeinschaft.

\end{document}